DISCOVERY OF THE EXPANSION OF THE UNIVERSE

Sidney van den Bergh
Dominion Astrophysical Observatory, Herzberg Institute of Astrophysics, National Research Council of Canada, 5071 West Saanich Road, Victoria, British Columbia, V9E 2E7, Canada, sidney.vandenbergh@nrc.ca

The myth that the expansion of the Universe was discovered by Hubble was first propagated by Humason (1931). The true nature of this discovery turns out to have been both more complex and more interesting.

On the last observing night before his retirement Milton Humason was tasked with teaching me how to use the Palomar 48-inch Schmidt telescope. By midnight Milt went to bed, apparently believing that his pupil was doing well enough to be left alone to use "The Big S". During the preceding three decades Humason had spent many very long nights collecting the spectra of faint galaxies that would enable Hubble & Humason (1931), using the new radial velocity observations by Humason (1931), to establish the linear expansion of the Universe beyond reasonable doubt. This work strengthened and confirmed the first hints of this expansion collected by Wirtz (1924), Lundmark (1925), Lemaître (1927), Hubble (1929) and de Sitter (1930). In his memoirs, collected during an oral history project [http:///www.aip.org/history/ohilist/4686.html] undertaken by the American Institute of Physics (Shapiro 1965), Milt Humason is quoted as follows: "The velocity-distance relationship started after one of the IAU meetings, I think it was in Holland. [The Third IAU meeting took place in Leiden in the Netherlands, September 5 - 13 1928.] And Dr. Hubble came home rather excited about the fact that two or three scientists over there, astronomers, had suggested that the fainter the nebulae were the more distant they were and the larger the red shifts would be. And he talked to me and asked me if I would try and check that out." Among the astronomers present at the IAU meeting, who might have been interested in a possible velocity-distance relationship for galaxies, were de Sitter, Hubble, Lemaître, Lundmark, Shapley and Smart.

The first tentative steps toward the discovery of the velocity-distance relationship were made by Wirtz (1922, 1924) and Lundmark (1925). In his 1922 paper Wirtz concludes that either the nearest or the most massive galaxies have the lowest redshifts. From the more extensive observational material available in 1924 Wirtz found that the radial velocities of spiral nebulae grow quite significantly with increasing distance. He was aware of the fact that the General Theory of Relativity predicted that redshifts should increase with increasing distance. Wirtz published his results in the Astronomische Nachrichten, the leading German astronomy journal. [Hubble received an A in his high-school German course (Christianson 1995, p. 31 and he also read German text books on corporate law Chistianson 1995, p. 79- so he would have had no trouble reading Wirtz's papers.] In 1925 Lundmark wrote "A rather definite correlation is shown between apparent dimensions and radial velocity, in the sense that the smaller and presumably more distant spirals have the higher space-velocity." In interpreting this result Lundmark opines that the observed Doppler shifts might be " ... effects consequent to the general theory of relativity." Lundmark's 1925 paper was published in the prestigious Monthly Notices of the Royal Astronomical Society, and was cited in Hubble (1929). However, Hubble dismisses this important paper with the comment that Lundmark's favored solution "offered little advantage."

Lemaître (1927) published a crucial paper that both established the expansion of the Universe and interpreted it as a consequence of the General Theory of Relativity. However, it is possible that Hubble (1929) was unaware of Lemaître's 1927 paper on the expansion of the universe because it had been published in French in a rather obscure publication. Nevertheless it is puzzling that Hubble and Lemaître would not have discussed this problem when they were both attending the 1928 IAU meeting in Holland. That Hubble dismissed Lundmark's paper might well be related to a bitter personal feud that resulted from the fact that Hubble (1926) had accused Lundmark of plagiarizing his system of galaxy classification. Hubble's failure to take Lundmark's 1925 paper seriously may have been largely responsible for the fact that this work has been ignored by many recent reviews of the discovery of the expansion of the universe (Block 2011, Kragh 1987, Luminet 1997, Shaviv 2011) - but not by Nussbaumer & Bieri (2009). In this connection it is of interest to note that Christianson (1995, p.230), in her biography of Hubble, writes about "[A] bitter, if little known, flare-up with Willem de Sitter, whom Grace [Hubble] had credited with spurring him to persue the velocity-distance relationship with the great telescopes at his command." She also writes that "Hubble became enraged at de Sitter's casual statement that several astronomers had commented on the [velocity-distance] relation." For a more detailed account of this matter [based on a letter from Hubble to de Sitter (dated August 21 1930) which is preserved in the Huntington Library - see Nussbaumer & Bieri (2009, p. 130).]

   In their 1931 paper Hubble and Humason berate the previous work by de Sitter (1930) because "[H]e arrived at the same numerical result" [see Table 1]. This was to be expected since he used essentially the same data. Exactly the same criticism may be applied to the paper by Hubble (1929), which used essentially the same input data that had previously been used by Lemaître (1927).

In summary the history of the discovery of the expansion of the Universe may be summarized as follows:

1922 From radial velocities of only 29 spirals Wirtz concludes that either the nearest or the most massive galaxies have the smallest redshifts.
1924 Using observations of 42 galaxies Wirtz (1924) concludes (my translation) "that there remains no doubt that the positive radial velocities of spiral nebulae grow quite significantly with increasing distance."
1925 Lundmark notes that the redshifts of small (presumably distant) spiral galaxies are larger than those of larger nearby ones.
1927 Lemaître derives the expansion rate of the Universe and explains its expansion in terms of the General Theory of Relativity.
1929 Hubble repeats Lemaître's work with essentially the same data and obtains similar results.
1930 de Sitter re-discusses mostly the same data more thoroughly and again finds the same result.
1931 Hubble & Humason obtain 40 new radial velocities which extend the determination of redshifts to the Leo cluster at a redshift of 19600 km/s. This places the reality of a linear velocity-distance relationship for galaxies beyond reasonable doubt.

The myth that Hubble discovered the velocity-distance relation seems to have originated with Humason (1931) [Who was at that time acting as Hubble's observing assistant]. Humason started his 1931 paper with the words "In 1929 Hubble found a relation connecting the velocities and the distances of the extragalactic nebulae for which spectra were then available." This point of view is supported by Hubble's statement (Christianson 1995, p.230) that "I consider the velocity-distance relation, its formulation, testing and confirmation, as a Mount Wilson contribution and I am deeply concerned in its recognition as such." By contrast Lemaître (1950) also wished to leave no doubt about who really discovered the expansion of the Universe. He writes about his 1927 paper: "[J']'y calcule le coefficient d'expansion (575 km par mégaparsecs, 625 avec une correction statistique contestable)…. Le titre de ma note ne laise aucun doubte sur mes intentions: <<Un Univers de masse constant et de rayon croissant rendant compte de la vitesse radiale des nébuleuses extra-galactiques.>>" The story told above shows that the actual history of the discovery of the expansion of the Universe was more complex and interesting than Humason suggested in 1931.

I thank Vivien Reuter for providing me with a listing of the "delegates" who attended the 1928 IAU meeting in Leiden. Also I am deeply indebted to numerous colleagues who have sent me e-mail messages related to various aspects of my recent article on Lemaître in this Journal. I am particularly indebted to Harry Nussbaumer, Jean-Pierre Luminet and David Block for their comments and encouragement. Thanks are also due to our librarian Bonnie Bullock for providing me with many ancient books and journal articles.

Table 1. Early determinations of the Hubble parameter

| Reference | $H_o$(km/s/Mpc) |
| --- | --- |
| Lemaître (1927) | 625 (weighted) |
| Lemaître (1927) | 575 (unweighted) |
| Hubble (1929) | 530 |
| de Sitter (1930) | 460 |
| Hubble & Humason (1931) | 558 |